\documentclass[letterpaper, 10 pt, conference]{ieeeconf}  

\IEEEoverridecommandlockouts                              

\usepackage{amsmath,amssymb,amsfonts}
\usepackage{graphicx}
\usepackage{bm}

\usepackage{algorithm}

\usepackage{booktabs}

\usepackage{textcomp}
\usepackage{amssymb}
\usepackage{amsmath}
\usepackage{graphicx}

\usepackage{xcolor}
\usepackage{makecell}
\usepackage{comment}
\usepackage{todonotes}

\usepackage{cite}
\usepackage{graphicx}

\usepackage{xcolor}
\definecolor{light-blue}{rgb}{0.3,0.5,0.8}
\usepackage[colorlinks=true, linkcolor=cyan, citecolor=cyan, filecolor=magenta, urlcolor=cyan]{hyperref}

\usepackage{caption}
\usepackage{subcaption}
\usepackage[capitalize]{cleveref}
\usepackage{url}

\usepackage{algpseudocode}

\title{\LARGE \bf
Cooperative Detour Planning for Dual-Task Drone Fleets
}

\author{Pengbo Zhu$^{1}$, Meng Xu$^{2}$, Andreas A. Malikopoulos$^{3}$, and Nikolas Geroliminis$^{2}$
\thanks{This research was supported in part by NSF under Grants CNS-2401007, CMMI-2348381, IIS-2415478, in part by MathWorks, and in part by Swiss National Science Foundation (SNSF) through project P500-2\_235379 and as a part of NCCR Automation, a National Centre of Competence in Research, funded by SNSF under Grant 51NF40\_225155). }
\thanks{$^{1}$Pengbo Zhu is with the School of Civil \& Environmental Engineering, Cornell University, Ithaca, NY, USA. (email: \texttt{pz283@cornell.edu})}%
\thanks{$^{2}$ Meng Xu and Nikolas Geroliminis are with Urban Transport Systems Laboratory, \'Ecole Polytechnique F\'ed\'erale de Lausanne, 1015 Lausanne, Switzerland,  (email: \texttt{meng.xu@epfl.ch,  nikolas.geroliminis@epfl.ch})}
\thanks{$^{3}$ Andreas A. Malikopoulos is with the Applied Mathematics, Systems Engineering, Mechanical Engineering, Electrical \& Computer Engineering, and School of Civil \& Environmental Engineering, Cornell University, Ithaca, NY, USA. (email: \texttt{amaliko@cornell.edu})}}
\begin{document}

\maketitle
\thispagestyle{empty}
\pagestyle{empty}

\begin{abstract}

As Urban air mobility scales, commercial drone fleets offer a compelling, yet underexplored opportunity to function as mobile sensor networks for real-time urban traffic monitoring. In this paper, we propose a decentralized framework that enables drone fleets to simultaneously execute delivery tasks and observe network traffic conditions. We model the urban environment with dynamic information values associated with road segments, which accumulate traffic condition uncertainty over time and are reset upon drone visitation. This problem is formulated as a mixed-integer linear programming problem where drones maximize the traffic information reward while respecting the maximum detour for each delivery and the battery budget of each drone. Unlike centralized approaches that are computationally heavy for large fleets, our method focuses on dynamic local clustering. When drones enter communication range, they exchange their belief in traffic status and transition from isolated path planning to a local joint optimization mode, resolving coupled constraints to obtain replanned paths for each drone, respectively. Simulation results built on the real city network of Barcelona, Spain, demonstrate that, compared to a shortest-path policy that ignores the traffic monitoring task, our proposed method better utilizes the battery and detour budget to explore the city area and obtain adequate traffic information; and, thanks to its decentralized manner, this ``meet-and-merge" strategy achieves near-global optimality in network coverage with significantly reduced computation overhead compared to the centralized baseline.

\end{abstract}


\section{Introduction}
\subsection{Background: Urban Air Mobility}
Urban Air Mobility (UAM) envisions the use of unmanned aerial vehicles (UAVs) and other low-altitude aircraft to support transportation and logistics in dense urban environments \cite{straubinger2020overview}. Among the various UAM applications, delivery drones have attracted particular attention for last-mile operations, where they can shorten delivery times, reduce operational costs, and complement ground-based logistics networks \cite{li2023drone}. With the maturing of regulatory frameworks such as the European civil aviation concept ``U-Space'' \cite{huttunen2019u} and the Federal Aviation Administration's Unmanned Aircraft System Traffic Management initiative ``Beyond''\cite{faa_beyond}, routine commercial drone delivery in populated areas is becoming increasingly feasible.

At the same time, drones offer unique advantages as mobile monitoring platforms. Unlike fixed roadside sensors, they can flexibly reposition in space, rapidly cover congestion hotspots, and collect aerial traffic data over wide urban areas. Large-scale aerial experiments, such as the pNEUMA campaign \cite{barmpounakis2020new}, have demonstrated that drone footage can provide rich, network-level traffic information that is difficult to obtain from sparse ground-based infrastructure alone. Subsequent work has further shown that coordinated drone fleets can actively improve traffic observability across urban road networks \cite{espadaler2025accurate}.

These two capabilities, delivery and monitoring, suggest an important opportunity. Delivery drones that routinely traverse city airspace need not be viewed as single-purpose logistics agents; when properly routed, they can simultaneously contribute to traffic monitoring and network observability while completing their primary delivery missions. This dual use is especially attractive when flight resources are limited, and monitoring needs are spatially and temporally dynamic. However, realizing it requires routing and coordination methods that explicitly balance delivery efficiency against the value of traffic information collected en route, rather than treating monitoring as an incidental by-product of flight. This trade-off, and the algorithmic challenges it poses, motivate the present work.

\subsection{Related Work}

\subsubsection{Routing and Path Planning for Delivery Drones}
The routing of delivery drones is rooted in the broader family of vehicle routing problems (VRPs), where the objective is to determine efficient routes for serving spatially distributed customers under operational constraints \cite{toth2014vehicle}. Early studies extended classical VRP formulations to aerial operations, giving rise to problems like the Flying Sidekick traveling salesman problem \cite{murray2015flying} and the VRP with drones  \cite{wang2017vehicle}. These foundational models are typically cast as mixed-integer linear programs (MILPs) that address path feasibility, delivery assignment, and minimization of travel distance or task completion time \cite{agatz2018optimization}. As the field of UAV systems grew and developed, the literature expanded into a rich family of variants incorporating the distinctive operational features of UAV systems.

Since then, numerous variants have been proposed to capture diverse operational constraints, including energy and battery limitations \cite{dorling2016vehicle}, delivery time windows \cite{jeong2019truck}, no-fly zones \cite{jeong2019truck}, and multi-trip dispatching with recharging \cite{jiang2025multi}. Despite the breadth of constraints considered, none of these variants treats the monitoring value of the route as a factor in path selection: the route is optimized solely for logistics performance.

In practice, multi-drone delivery operations also face communication constraints: drones can only exchange information with peers within a limited wireless range, and the communication network topology changes continuously as vehicles move \cite{chung2018survey}. This makes centralized coordination, where a single controller batches all requests and computes globally optimal routes, difficult to sustain in real time, especially as fleet size grows. At the same time, the operational environment is inherently stochastic and dynamic: delivery requests arrive sequentially, and the information available to each drone is local and incomplete. Our work builds directly on this motivation: we propose a dynamic local clustering mechanism in which drones within communication range jointly re-optimize their routes in real time, without requiring global information or centralized batch processing.

\subsubsection{Drone-Based Traffic Monitoring}
A separate body of work has investigated drones as tools for traffic monitoring. This literature emphasizes spatial coverage, observation scheduling, and the information value of aerial data, rather than delivery optimization. Drone-based monitoring has been applied to the ground-vehicle trajectory collection \cite{krajewski2018highd}, traffic state estimation \cite{ke2016real}, and anomaly detection \cite{tran2023uit}, particularly where the fixed sensor infrastructure, such as loop detectors, is sparse or absent.

An important distinction in this literature is between \emph{dedicated} and \emph{opportunistic} monitoring. Works in drone monitoring typically assume that drones are deployed as dedicated sensing agents whose sole mission is data collection \cite{butilua2022urban}. Their results motivate a natural follow-up question: can delivery drones, already airborne for logistics purposes, serve as opportunistic contributors to urban traffic monitoring, and if so, how should their routes be designed to make this dual use effective? Therefore, despite the maturity of drone-based monitoring as a standalone application, the literature has overlooked the scenario in which monitoring is performed by drones that are simultaneously carrying out delivery missions and are therefore subject to delivery-oriented constraints such as maximum allowable detour and limited battery budget.

\subsubsection{Dual-Function Drone Operations}
A limited but growing body of literature has begun to jointly consider delivery and monitoring within a single UAV framework. The most common approach formulates this integration as a VRP-type extension, where monitoring is abstracted as servicing a set of pre-specified task locations or points of interest (POIs). For example, Rottondi et al. \cite{rottondi2021scheduling} treat monitoring as node-visit requirements within a multi-service VRP, while Chen et al. \cite{chen2024ddl} augment delivery routing with hovering-time allocation at monitoring POIs. Along a related direction, Sun et al. \cite{sun2026robust} propose a collaborative truck--drone framework for post-disaster operations in which drones execute both delivery and monitoring at discrete locations. Despite differences in mathematical structure, these works largely share an offline, static planning premise and a node-centric abstraction of monitoring. This perspective differs from road-network-based monitoring, where the objective is to maintain up-to-date information over road segments and requires segment-level information refresh.
 
Beyond VRP-based formulations, Xiang et al. \cite{xiang2021reusing} study the reuse of delivery drones for urban monitoring by selecting among pre-specified candidate routes and allocating hovering effort at discrete sensing locations under energy and deadline constraints. Similarly, Gao et al. \cite{gao2024sharing} investigate sharing instant-delivery UAVs for monitoring fixed POIs and stochastic events. In all these cases, monitoring objectives are represented as visiting discrete locations. 
 
Across these studies, several gaps persist that our work aims to address: (i) monitoring is almost universally abstracted as discrete node/POI visits rather than fine-grained, road-segment-level monitoring; (ii) decentralized, real-time replanning under limited communication has received little attention in this integrated setting. Addressing this combination of challenges is the focus of this study.

\subsection{Contributions}

The main contribution of this work is a novel detour planning algorithm for dual-task delivery drones, where each drone completes its delivery task while respecting its budget constraints and, at the same time, monitors traffic conditions. 
Unlike centralized control algorithms, we propose a ``meet-and-merge'' decentralized scheme that performs dynamic local clustering, allowing drones to coordinate with neighboring drones. Joint optimization for multiple drones in the same cluster is performed. This decentralized structure makes the proposed method both computationally scalable and practical under real-world communication limitations.

The remainder of this paper is organized as follows: 
In \cref{sec: Motivation}, the motivation for the cooperative planning problem for delivery drones is provided. 
In \cref{sec: Traffic Information Reward}, we model the traffic knowledge gain as a linear model considering the traffic information density and the length of each road segment. 
We then formulate the objective function as maximizing the jointly collected traffic knowledge for a multi-drone cluster as a mixed-integer linear programming problem in \cref{sec: MILP}. 
The proposed algorithm is tested on an agent-based simulator using the real city road network of Barcelona, Spain, in \cref{sec: Case Study}. 
Finally, we conclude the work and present future research directions in \cref{sec: Conclusion}.
\section{Problem Formulation}
\subsection{Motivation}\label{sec: Motivation}
Commercial drones mounted with cameras are able to provide a bird’s-eye view of traffic conditions while carrying out delivery tasks and flying over urban areas. In this context, drones can accomplish two tasks: 1) parcel delivery and 2) traffic monitoring, by following a single cooperative route. 

For the purpose of collecting traffic information on road segments, we assume all drones will fly following the urban road network, which is modeled as a directed graph $\mathcal{G} = (\mathcal{V}, \mathcal{E}, \omega)$, where $\mathcal{V}$ represents the set of intersections, $\mathcal{E}$ represents the set of road segments, and $\omega$ is the length of the road segment.

\begin{figure}[htb]
\vspace{3mm}
\centering
\begin{subfigure}[ht]{0.48\textwidth}
\includegraphics[width=\textwidth]{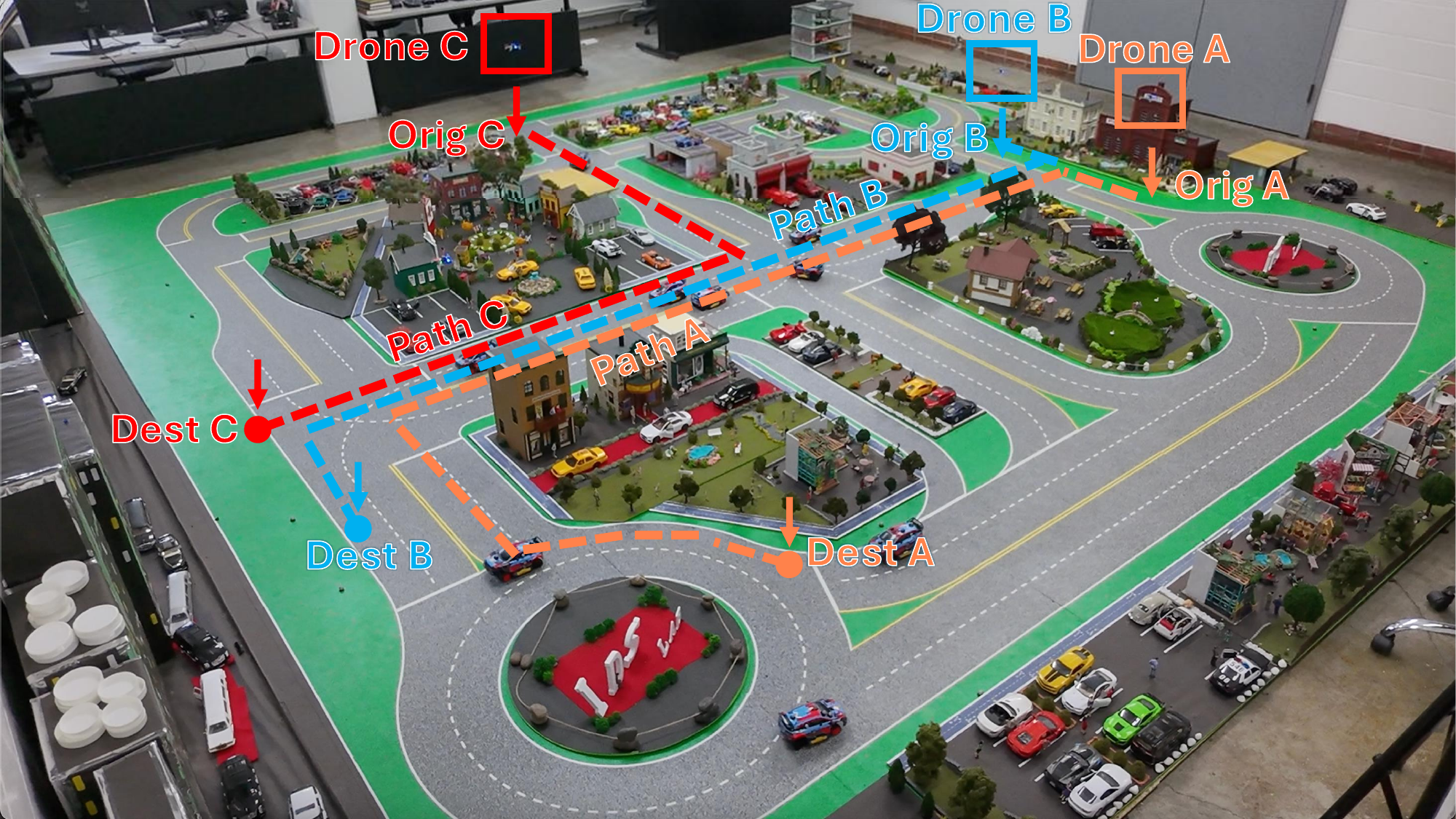}
 \caption{Shortest Paths}\label{fig:shortest}
\end{subfigure}\\
\begin{subfigure}[ht]{0.48\textwidth}
\includegraphics[width=\textwidth]{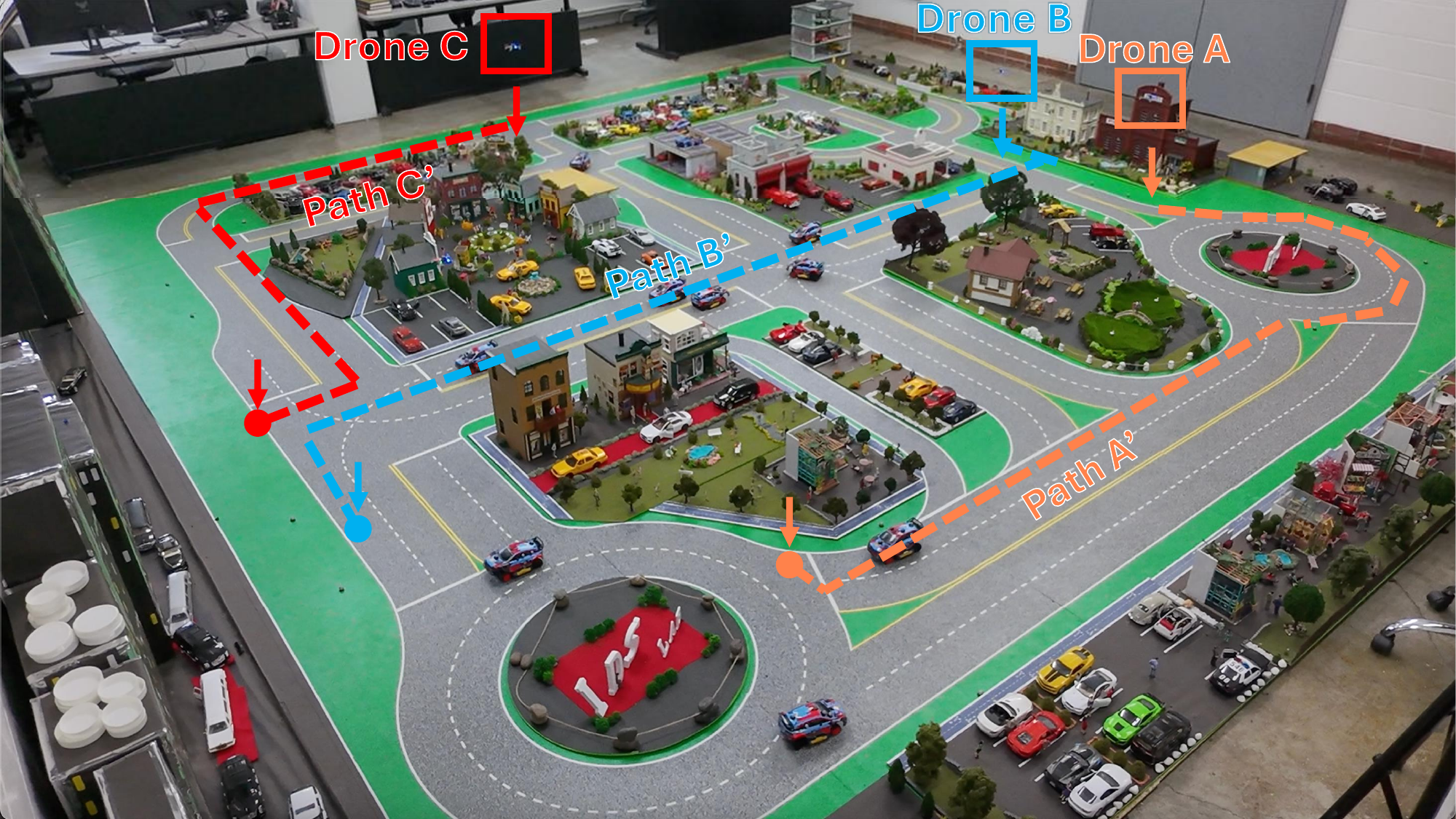}
 \caption{Cooperative Detour Paths}
\label{fig:detour}
\end{subfigure}
\caption{The testing scenario with 3 drones operating in a 1:25 scaled smart city testbed (IDS$^3$C, see \cite{chalaki2021CSM, remer2019multi} for more information). The demo video is available on \url{https://sites.google.com/view/cdc2025-drone-with-a-mission/home}.}
    \label{fig:ScaledCity_overview}
\end{figure}

\vspace{-0.75cm}

As illustrated in \cref{fig:shortest}, when a drone $k$ is assigned a logistics task, it picks up the parcel at its origin $O_k$ and delivers it to its destination $D_k$ following a path determined by the operational strategy. If Drone A and Drone B follow the shortest path in terms of travel distance to deliver their parcels, they will visit the exact same road segments within a very short time. Similarly, Drone C will also visit the same intersection in the middle of the scaled city. However, with efficient path planning for this cluster of multiple drones as illustrated in \cref{fig:detour}, the agents differentiate their paths to cover a larger urban area, obtaining better knowledge of the traffic information. Even though Drones A and C take longer paths involving detours, this approach effectively balances delivery efficiency with the traffic surveillance task.

This example demonstrates that, in a dual-objective context for joint logistics and traffic monitoring, the shortest path is not necessarily the optimal choice; instead, an efficient path-planning algorithm for a multi-drone fleet is essential to the system's operation. Rather than solving the route planning for all drones, which is computationally heavy, in this study, our proposed method is implemented within a cluster of drones that are within local communication range, where they share a synchronized traffic knowledge belief matrix and jointly plan their paths to their respective destinations. 

In the following subsections, we first model the traffic information reward considering the last visit timestamp, and subsequently maximize the total cooperative information collected by the drone cluster, subject to battery and detour constraints.

\subsection{Traffic Information Reward}\label{sec: Traffic Information Reward}

The knowledge of the traffic status for each edge $(i,j)$ is affected by its uncertainty growth rate (i.e., how fast the traffic status changes on this road) and its monitoring history (i.e., when the road segment was last visited). The information reward $R_{ij}$ is 
\begin{equation}
    R_{ij} = \min \left( \beta_{ij}\omega_{ij} (t^{scan}_{ij} - T^{last}_{ij}), R_{ij}^{max} \right),
\end{equation}
where $T^{last}_{ij}$ represents the true global timestamp when edge $(i,j)$ was last visited, and $t^{scan}_{ij} \in \mathbb{R}_{\geq 0}$ is the continuous decision variable representing the planned effective scan time.

However, in a decentralized multi-agent system, the true global monitoring history $T^{last}_{ij}$ is partially observable. Therefore, during the cooperative replanning phase, the drone cluster relies on its synchronized belief matrix $\mathcal{B}$ as the local estimator for the environment's state. The MILP formulation substitutes the ground truth with the shared belief, evaluating the expected information gain using $\hat{T}^{last}_{ij} = \mathcal{B}_{ij}$. 

The total uncertainty growth rate for the entire edge is the uncertainty growth density $\beta_{ij}$ multiplied by the length of the edge $\omega_{ij}$. This expected information gain is bounded by a saturation limit $R_{ij}^{max}$, which prevents infinite reward accumulation for unvisited edges.

\subsection{MILP for cooperative detour planning}\label{sec: MILP}

We consider a cluster of drones $\mathcal{K}$ that have established local communication. 
For each drone $k \in \mathcal{K}$, we define a binary decision variable $x_{ijk}$ as:
\begin{equation}
    x_{ijk} = 
    \begin{cases} 
    1, & \text{if drone } k \text{ traverses edge } (i,j) \in \mathcal{E}, \\
    0, & \text{otherwise}.
    \end{cases}
\end{equation}

The cooperative edge coverage variable $y_{ij}$ is defined as:
\begin{equation}
    y_{ij} = 
    \begin{cases} 
    1, & \text{if any drone } k \in \mathcal{K} \text{ traverses edge } (i,j) \in \mathcal{E}, \\
    0, & \text{otherwise}.
    \end{cases}
\end{equation}

Finally, $t_{i,k} \geq 0$ is a continuous variable representing the relative arrival time of drone $k$ at node $i$.

The cooperative information path planning (IPP) problem is formulated as an asynchronous joint optimization subject to detour budget and battery constraints. The objective is to maximize the total time-dependent information gain collected by drones within the local cluster, given by:
\begin{subequations}\label{MILP}
\begin{align}
    & \text{Maximize} \quad J = \sum_{(i,j) \in \mathcal{E}} R_{ij}, \nonumber \\
    & \text{subject to} \nonumber \\
    & \sum_{j \in \delta^+(O'_k)} x_{O'_k j k} - \sum_{j \in \delta^-(O'_k)} x_{j O'_k k} = 1, \quad \forall k \in \mathcal{K}, \label{cons: Start} \\ 
    & \sum_{j \in \delta^+(D_k)} x_{D_k j k} - \sum_{j \in \delta^-(D_k)} x_{j D_k k} = -1 \quad \forall k \in \mathcal{K}, \label{cons: End} \\ 
    & \sum_{j \in \delta^+(i)} x_{ijk} - \sum_{j \in \delta^-(i)} x_{jik} = 0 \quad \forall k \in \mathcal{K}, \forall i \in \mathcal{V} \setminus \{O'_k, D_k\}, \label{cons: Flowcontinuity} \\ 
    & t_{O'_k, k} = \Delta t_k, \quad \forall k \in \mathcal{K}, \label{cons: Timeoffset} \\ 
    & t_{j, k} \ge t_{i, k} + \tau_s(i,j) - M(1 - x_{ijk}) \quad \forall k \in \mathcal{K}, \forall (i,j) \in \mathcal{E}, \label{cons: Finalarrival} \\
    & t_{D_k, k} \le B_{eff,k} + \Delta t_k, \quad \forall k \in \mathcal{K}, \label{cons: Budget} \\
    & \sum_{k \in \mathcal{K}} x_{ijk} \ge y_{ij}, \quad \forall (i,j) \in \mathcal{E}, \label{cons: Cooperative} \\ 
    & t^{scan}_{ij} \le t_{j,k} + M(1 - x_{ijk}), \quad \forall k \in \mathcal{K}, \forall (i,j) \in \mathcal{E}, \label{cons: Reward} \\ 
    & R_{ij} \le \beta_{ij}\omega_{ij} \left( T_{curr} + t^{scan}_{ij} - \mathcal{B}_{ij} \right), \quad \forall (i,j) \in \mathcal{E}, \label{cons: InfoReward} \\ 
    & R_{ij} \le R_{max, ij} \cdot y_{ij}, \quad \forall (i,j) \in \mathcal{E}, \label{cons: Rewardbound}
\end{align}
\end{subequations}
where $\tau_{s}(i,j)$ is the shortest travel time of the edge $(i,j)$, calculated as the length of the edge divided by the flying speed of the drone.
To ensure continuous-time physical feasibility during discrete-time route planning, let $T_{curr}$ be the global time when a new event triggers the planning cycle (e.g., the start of one parcel delivery, or the appearance of a new drone in the current cluster). At time $T_{curr}$, drone $k$ is actively executing its previously planned trajectory. We assume it must finish traversing its current road segment. Let the drone be traveling along edge $(u_k, v_k)$ at time $T_{curr}$. We define $v_k$ as the new planning origin $O'_k$ for the upcoming optimization cycle: $O'_k = v_k$. Constraints \eqref{cons: Start} and \eqref{cons: End} guarantee that the route begins at node $O'_k$ and ends at node $D_k$ (the unchanged delivery destination of drone $k$). Constraint \eqref{cons: Flowcontinuity} ensures flow continuity at all nodes except the start and end nodes, maintaining a continuous path. 

The drone will arrive at $O'_k$ at some time $t_{arr}(v_k) \geq T_{curr}$. We calculate the exact time offset $\Delta t_k$, representing the mandatory period before the new joint plan can actually commence for drone $k$ as $\Delta t_k = t_{arr}(v_k) - T_{curr}$.

Instead of starting at $t=0$, the arrival time variable at the new origin is initialized to the offset, as shown in Constraint \eqref{cons: Timeoffset}. Constraint \eqref{cons: Finalarrival} governs the temporal evolution of the system. If edge $(i,j)$ is traversed, the arrival time at $j$ is strictly linked to the arrival time at $i$.

$B_{eff, k}$ denotes the effective remaining travel budget for drone $k$ (unit: seconds). This is calculated as the minimum of the remaining battery budget and the maximum allowable delivery detour delay:
\begin{equation}
    B_{eff, k} = \min \left( (1 + \alpha)\tau_{s}(O_k, D_k) - (T_{curr} - T_{O_k,k}), \ B_{b, k} \right),
\end{equation}
where $\alpha \geq 0$ is the allowed detour factor, $\tau_{s}(O_k, D_k)$ is the theoretical shortest path time from the initial pickup to the destination, and $T_{O_k,k}$ is the time the parcel was originally picked up.
The remaining battery budget $B_{b, k}$ is computed by dividing the drone energy $E_k$ by the average energy consumption rate $c_e$ to get the remaining flight time.
Constraint \eqref{cons: Budget} bounds the final arrival time at the destination, that the drone should retain sufficient budget to reach its global destination. 

Constraints \eqref{cons: Cooperative} and \eqref{cons: Reward} govern the earliest scan time among all cooperative drones in the cluster to prevent redundant reward accumulation. Finally, Constraint \eqref{cons: InfoReward} computes the reward based on the age of information bridging the global simulation time $T_{curr}$ and the relative planned scan time $t^{scan}_{ij}$, bounded by a saturation limit as shown in Constraint \eqref{cons: Rewardbound}.

Solving this MILP problem provides the optimal set of edges, which maximizes the total collected traffic information while adhering to constraints. The route of each drone $k$ is then constructed by sequentially connecting these selected edges $x^*_{ijk}$, forming a continuous path from their current path start node $O'_k$ to the destination $D^k$.

\subsection{Decentralized Event-Triggered Cooperative Detour Path Planning} \label{sec: CIPP}

To implement the proposed method in real-time, we propose a decentralized, event-triggered algorithm equipped with edge pruning. 

\subsubsection{Event-Trigger and Traffic Information Synchronization}
When a drone picks up a parcel, it calculates an initial path by solving \eqref{MILP} as a single agent, respecting its maximum detour and battery budgets. In this decentralized multi-agent system, rather than relying on a central controller, each drone maintains its own local map of the world—more specifically, a matrix $\mathcal{B}$ that records its local belief regarding the last visit time of all edges in the network. 

As drones fly their assigned parcels to their destinations, they continuously broadcast their locations within their proximity. If two or more drones encounter each other within a communication radius $R_c$, they form a cooperative cluster $\mathcal{K}$. They then synchronize their traffic knowledge by taking the element-wise maximum of their individual belief matrices:
\begin{equation}
    \mathcal{B}_{ij} = \max_{k \in \mathcal{K}} \mathcal{B}_{k, ij} \quad \forall (i,j) \in \mathcal{E}.
\end{equation}

Leveraging the shared belief matrix $\mathcal{B}$, the drones formulate and solve the MILP defined in \eqref{MILP}. The solver evaluates the cluster's combined state to identify high-value, long-unvisited road segments. 

\subsubsection{Edge Pruning}
In order to maintain computational tractability in real-time, an ellipsoid bounding technique is applied to the graph. The shared routing graph of the cluster is reduced such that it only contains nodes $v$ reachable by at least one drone $k \in \mathcal{K}$, given its effective remaining budget:
\begin{equation}
    d(O'_k, v) + d(v, D_k) \le B_{eff, k}.
\end{equation}

\subsubsection{Implementation of the algorithm}

The implementation is summarized in Algorithm \ref{alg:cooperative_ipp}.

\begin{algorithm}[!tbp]

\caption{Decentralized Event-Triggered Cooperative Detour Path Planning}
\label{alg:cooperative_ipp}
\begin{algorithmic}[1]
\State \textbf{Input} Urban road network $\mathcal{G}(\mathcal{V}, \mathcal{E},\omega)$, Local belief matrices $\mathcal{B}_k$, Global time $T_{curr}$
\State \textbf{Event Trigger:} Drones approach within communication range $R_{c}$ to form cluster $\mathcal{K}$.
\State \textbf{Belief Synchronization:} Form the shared belief matrix by adopting the most recent observation timestamps from all agents:
\Statex \qquad $\mathcal{B}_{ij} \gets \max_{k \in \mathcal{K}} \mathcal{B}_{k, ij} \quad \forall (i,j) \in \mathcal{E}$
\For{each drone $k \in \mathcal{K}$}
    \State Extract current trajectory edge $(u_k, v_k)$
    \State Set new planning origin: $O'_k \gets v_k$
    \State Calculate expected arrival offset: $\Delta t_k$
    \State Calculate effective remaining Budget $B_{eff, k}$
    \State Update local belief: $\mathcal{B}_k \gets \mathcal{B}$
\EndFor \\
Apply Ellipsoid Pruning to $\mathcal{G}$ using $B_{eff,k}$
\State $\mathbf{x}^*, \mathbf{t}^* \gets \text{SolveMILP}()$
\If{Solution is optimal}
    \For{each drone $k \in \mathcal{K}$}
        \State Extract optimal path $\mathcal{P}^*_k$ from $\mathbf{x}^*$
        \State Update drone trajectory to $\mathcal{P}^*_k$
    
    \EndFor
\Else
    \State \Return Default to shortest paths to prevent delivery failure
\EndIf
\end{algorithmic}
\end{algorithm}
\section{Case Study}\label{sec: Case Study}
\subsection{Numerical Experimental Setup}
We evaluate the proposed framework on a real-world urban road network from the Eixample district in Barcelona, Spain. The network consists of 805 intersection nodes and 1,397 road segments spanning approximately $7.6\,\text{km}^2$. Since UAVs operate above the road network and are not restricted by one-way regulations, all segments are treated as bidirectional, yielding a dense topology with an average node degree of 6.9. 
The traffic uncertainty growth density $\beta_{ij}$ is estimated from real traffic data related to the average traffic volume and the temporal variability of vehicle kilometers traveled, where a higher $\beta_{ij}$ value reflects a higher rate of traffic uncertainty accumulation.

A homogeneous fleet of UAVs is deployed, each flying at a constant speed 8.0 m/s. Energy is modeled via a battery system: initial charge levels are randomized between 30\% and 90\%; the consumption rate is $c_e = 1.68\% / \mathrm{min}$ (approximately one hour of flight when fully charged); and the charging rate is $\gamma_c = 4.8\%/ \mathrm{min}$ (full recharge in roughly 20 minutes). Two charging stations are available within the network. Initial UAV positions are generated randomly across the network. The maximum allowable detour factor is set as $\alpha = 30\%$ of the original shortest trip. And the communication range of drones is $R_c = 300$m.

\begin{figure}
    \centering
    \includegraphics[width=1\linewidth]{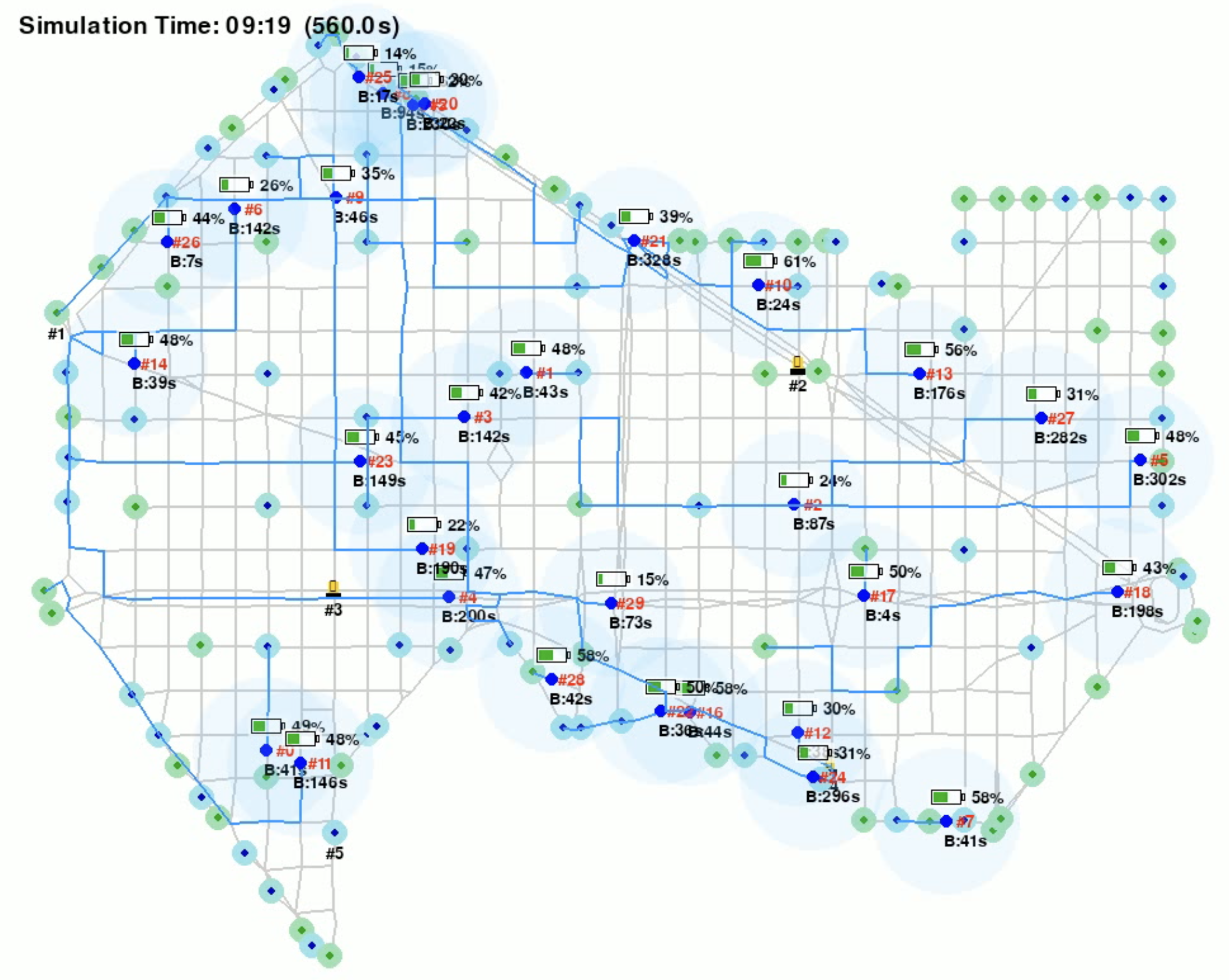}
    \caption{Snapshot of the Barcelona network with delivery drones whose communication range is shown as the blue circular disk. The origins of parcels are marked as green circles, while destinations are in blue. The color of the trajectories: a drone actively delivering a parcel is colored in blue, and the path is designed by our proposed method; a purple trajectory indicates that the drone is empty and en route to pick up a parcel; an orange trajectory indicates that the drone is returning to the closest charging station to charge its battery. The demo video of the proposed method is available on \url{https://sites.google.com/view/cdc2025-drone-with-a-mission/home}.}
    \label{fig:Barcelona_single}
    \vspace{-0.2cm}
\end{figure}

Delivery tasks arrive dynamically according to a Poisson process at a rate of 0.8 orders per second. We model OD pairs using calibrated human mobility patterns from Aimsun \cite{casas2010traffic}. This modular design allows seamless replacement with actual parcel data once available. For the sake of simplicity, once a drone finishes its last delivery task and becomes available, we assign it a new parcel that is closest to it and that it can complete within its battery budget. If the remaining budget is not sufficient to complete any delivery task, the drone will fly to the nearest charging station to recharge. \cref{fig:Barcelona_single} shows a snapshot of the simulator. 

\subsection{Result and Analysis}

To evaluate system efficiency, we consider the following key performance metrics: 1) \textbf{total system information gain}, defined as the sum of the reward collected over the entire simulation horizon; 2) \textbf{spatial coverage}, measured as the percentage of the total city edges visited at least once during the simulation; 3) \textbf{average age of information (AoI)}, which is the average time since the last visit for all edges in the city at the end of the simulation horizon; 4) \textbf{average parcel delivery delay}, representing the average percentage increase in delivery time compared to the strict shortest path; 5) \textbf{total MILP calls}, representing how many time the replanning scheme is triggered during the simulation horizon; 6) \textbf{average CPU time per MILP call}, which measures the computational latency of a single solver execution.

\begin{table*}[!htbp]
\vspace{3mm}
    \centering
    \caption{Performance Comparison of Path Planning Strategies}
    \label{tab:performance_metrics}
    \setlength{\tabcolsep}{3pt} 
    \begin{tabular}{lcccccc}
        \toprule
        \textbf{Method} & \textbf{Total info gain} & \textbf{Spatial coverage ($\%$)} & \textbf{Avg. AoI ($\%$)} & \textbf{Avg. delivery delay  ($\%$)} & \textbf{Total MILP Calls}& \textbf{Avg. CPU time ($s$)}\\
        \midrule
        Shortest & 350,214 & 46.92 & 68.71 & - & - &- \\
        Distributed & 711,301 & 76.54 & 50.30 & 21.46 & 175 &3.12\\
        Centralized & 831,278 & 84.10 & 46.30 & 25.11& 1655 &4.44\\
        \textbf{Decentralized (Proposed)} & 789,091 & 80.67 & 46.46 & 22.15 & 609  &3.99\\
        \bottomrule
    \end{tabular}
\end{table*}

For comparison, we evaluate our proposed method against three benchmark strategies. The first is the \textit{Shortest} strategy, where all drones lack communication and follow their shortest paths (computed using the Floyd–Warshall algorithm \cite{Floyd1962}) to deliver the parcels, regardless of the traffic monitoring objective. The second is the \textit{Distributed} strategy, in which all drones plan their paths according to their individual local beliefs, but they do not exchange information nor synchronize their belief matrices with one another. Finally, the \textit{Centralized} strategy considers the whole fleet as a single cluster (i.e., the communication range is infinite), keeping the shared belief matrix perfectly synchronized with the true global visit time matrix at all times.

\begin{figure}[!h]
    \centering
    \includegraphics[width=0.48\textwidth]{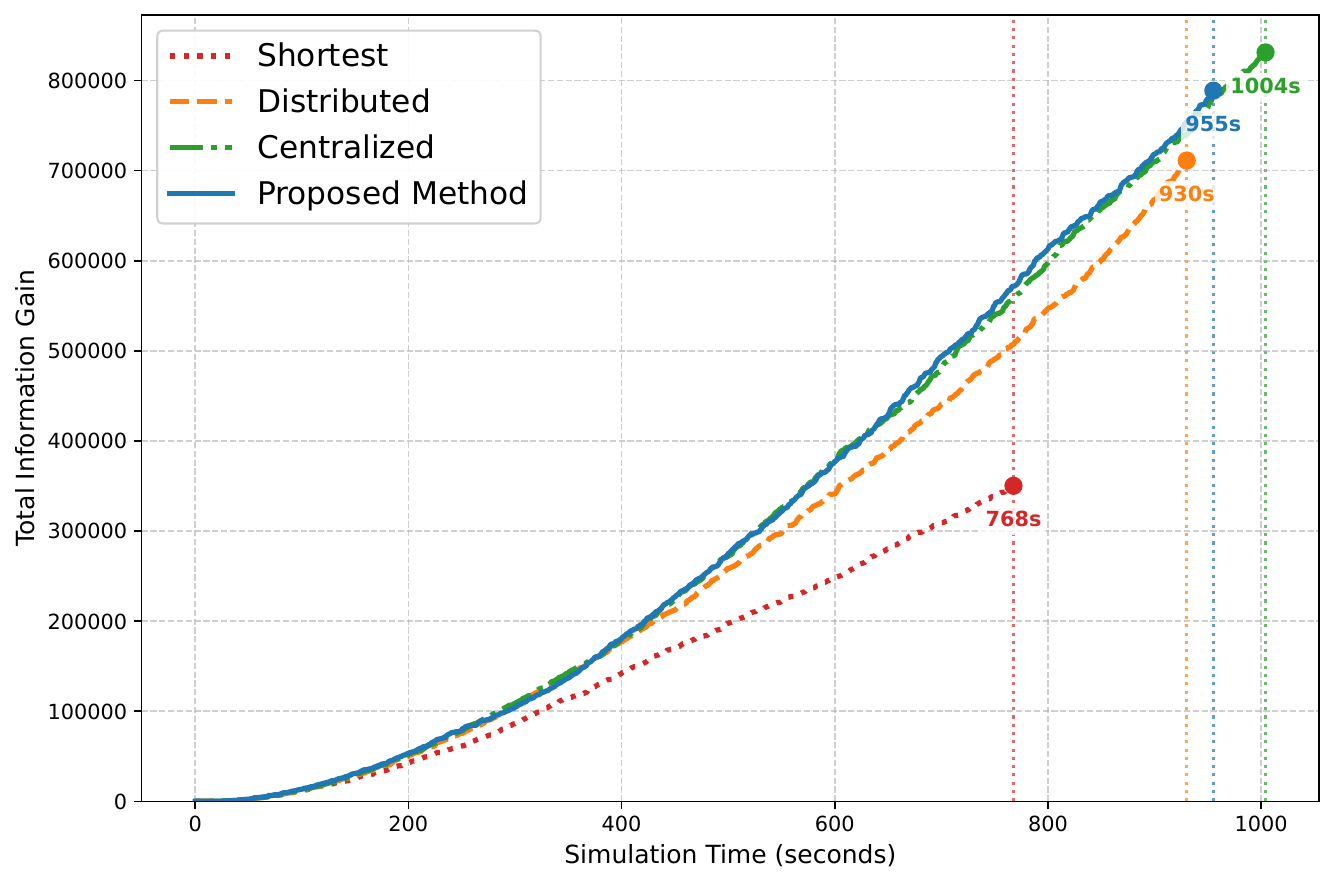}
    \caption{The evolution of total information gain over time for different planning strategies}
    \label{fig: InfoReward}
\end{figure}

Table \ref{tab:performance_metrics} presents the performance metrics of the four path planning strategies for a fleet of 30 drones delivering 150 parcels. As expected, the Shortest path baseline yields the lowest spatial coverage (46.92\%) and total information gain. By strictly minimizing delivery time and ignoring the traffic monitoring objective, drones repeatedly traverse the same central routes without exploring the broader network. In contrast, the proposed \textit{Decentralized} method successfully utilizes the allowable 30\% detour budget (incurring an average delay of 22.15\%) to actively explore the environment, significantly outperforming the \textit{Distributed} scheme in total information gain. This performance gap highlights the effectiveness of our designed ``meet-and-merge'' belief synchronization policy. Because the \textit{Distributed} baseline isolates each drone's belief state throughout its trip, agents frequently overestimate the available reward on previously visited edges, leading to redundant explorations. The proposed method resolves this by dynamically sharing information, achieving a higher spatial coverage (80.67\%) than the \textit{Distributed} approach. Furthermore, the normalized Average Age of Information (AoI), designed to evaluate the freshness of the collected data, shows that the proposed method (46.46\%) substantially improves data freshness over the \textit{Distributed} method and performs nearly on par with the \textit{Centralized} upper bound.

While the \textit{Centralized} baseline provides the theoretical bound for spatial coverage and information gain, it suffers from severe computational load. Because the \textit{Centralized} baseline assumes an infinite communication range, any single localized event, such as a drone picking up a new order, forces a globally coupled replanning of the entire active fleet. To ensure the MILP solver remained computationally tractable for routing all 30 drones simultaneously, these global updates had to be processed using batched optimization, solving the fleet in sequential chunks of up to three drones. Consequently, the \textit{Centralized} approach triggered 1,655 MILP calls, which is 2.7 times more than the proposed method with the highest average CPU time per call, leading to massive accumulated computational delays that make it impractical for reality. In contrast, the proposed method naturally bounds computational complexity by leveraging localized communication to form small, dynamic clusters. It resulted in a 63\% reduction in replanning calls compared to the \textit{Centralized} method, proving that our proposed method maintains highly competitive coverage and information gain while remaining highly scalable for real-time implementation.

In addition, as illustrated in \cref{fig: InfoReward}, which shows the accumulated knowledge gain over time for all methods, the \textit{Shortest} policy finishes all parcel delivery tasks the fastest but results in low traffic knowledge collection due to redundant visits to the same city areas. In contrast, our proposed method achieves almost the same level of knowledge as the \textit{Centralized} method, and a higher total reward than the \textit{Distributed} scheme. It is noticeable that the proposed and centralized methods have the steepest slopes compared to the distributed and shortest methods, indicating that their knowledge collection speed is faster.

\begin{figure}[htb]
\vspace{2mm}
\centering
\begin{subfigure}[ht]{0.45\textwidth}
\includegraphics[width=\textwidth]{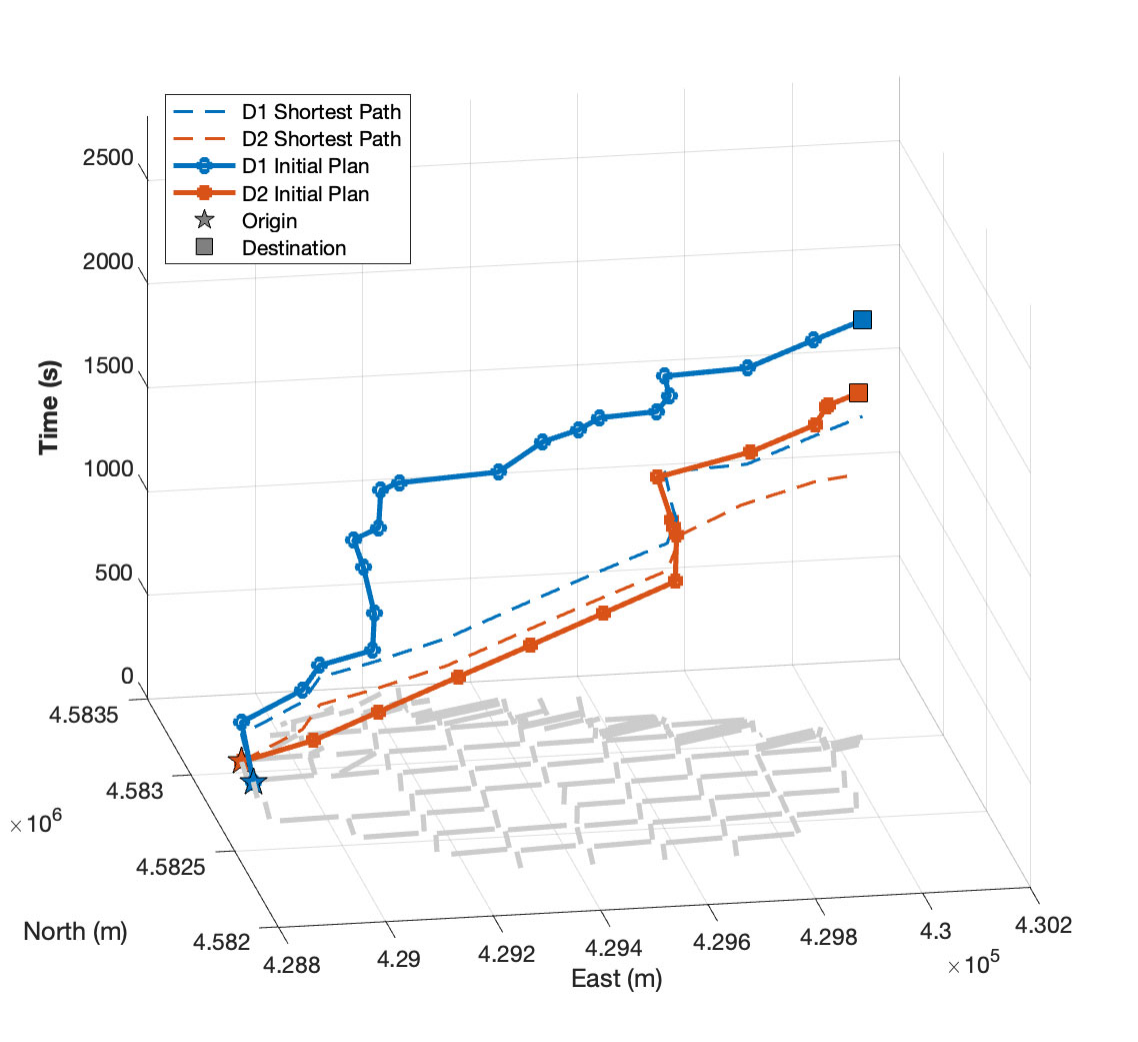}
 \caption{Initial Planning v.s. Shortest Paths}\label{fig:initial}
\end{subfigure}\\
\begin{subfigure}[ht]{0.45\textwidth}
\includegraphics[width=\textwidth]{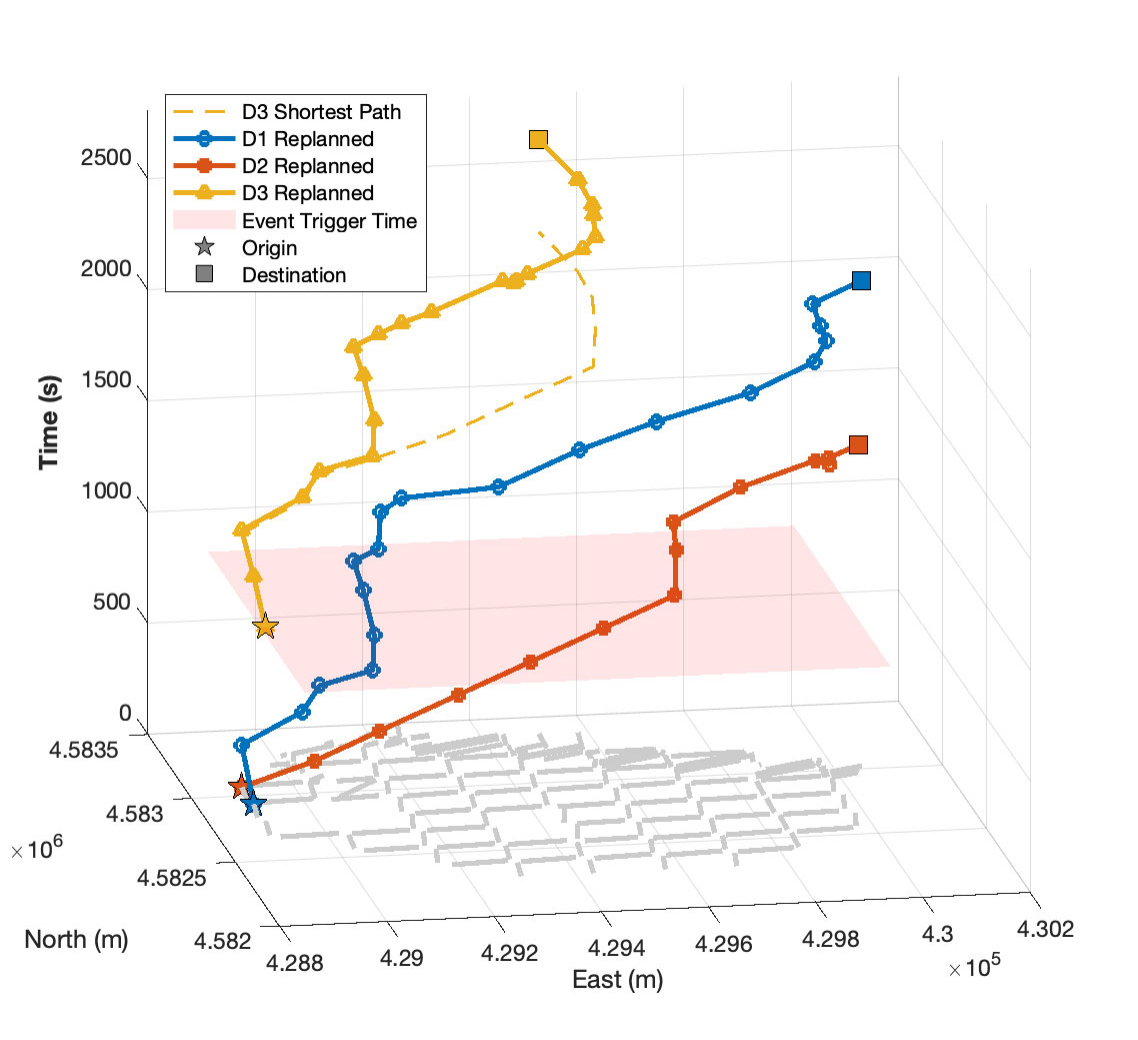}
 \caption{Replanning when Drone 3 joints}
\label{fig:replanning}
\end{subfigure}
\caption{Spatiotemporal cooperative routing of a multi-drone cluster}
\label{fig:MergeReplanningExample}
\vspace{-0.3cm} 
\end{figure}

To clearly visualize the proposed ``meet-and-merge'' behavior, \cref{fig:MergeReplanningExample} illustrates the spatiotemporal trajectories of a dynamically formed drone cluster. For visual clarity, the figure isolates three interacting drones and focuses on their specific localized region within the Barcelona road network. The baseline shortest paths are denoted by dashed lines, while the trajectories generated by the proposed cooperative routing algorithm are shown as solid lines. If the drones follow their shortest paths, their trajectories would exhibit severe spatiotemporal overlap. Consequently, the agents would repeatedly visit the same sequence of edges within a short time window, resulting in redundant observations and a lower collection of fresh traffic information. In contrast, the initial planned paths demonstrate that the proposed method actively forces the drones to detour and explore a broader set of edges, maximizing the aggregate information reward while satisfying individual flight budgets. When a third drone approaches the communication radius, they form a new cluster and trigger an event-based replanning phase, denoted by the red horizontal plane. Upon triggering, the proposed algorithm formulates a joint optimization problem for all three drones. As seen in the adjusted trajectories above the red plane, the paths are dynamically replanned. The updated routing successfully accommodates the introduction of the third drone, collaboratively redistributing the remaining exploration tasks to ensure optimal coverage of the road network.

\section{Conclusion}\label{sec: Conclusion}

In this paper, we addressed cooperative path planning for delivery drones in urban air mobility systems. In addition to performing logistics tasks, the drones are equipped with cameras, enabling them to monitor traffic conditions while flying over the urban road network. Our proposed framework explicitly considers this dual-task mission for multi-drone fleets and introduces a detour planning algorithm that aims to maximize traffic information gain along planned delivery routes. First, we model the traffic information state as a linear function that evolves; then, we formulate the cooperative path planning problem as an MILP that accounts for battery limitations and maximum allowable detours. To reduce the computational burden, we developed a decentralized scheme with dynamic clustering, in which drones with communication constraints exchange their beliefs about traffic conditions with neighboring drones. We compared the performance of the proposed method with several benchmark approaches and demonstrated its clear advantage in balancing delivery efficiency with traffic surveillance effectiveness.

The traffic information function is estimated from historical data; however, traffic conditions are strongly influenced by exogenous factors such as accidents and weather events, which can exhibit pronounced spatiotemporal variability. Future research should focus on integrating online learning of traffic conditions and belief updating. In addition, incorporating charging-scheduling decisions into the framework is a promising direction that is expected to further improve overall system efficiency. Finally, exploring aerial-ground operation by coordinating the drone fleet with connected and automated vehicles (CAVs) presents an exciting avenue for multimodal data fusion. To empirically validate these cooperative strategies, future work will also focus on interfacing the current simulator directly with the physical 1:25 scaled city testbed \cite{chalaki2021CSM} via an API, bridging the gap between digital simulation and real-world deployment.

\section*{Acknowledgment}
The authors would like to thank Weihao Sun and Zachary Wu for their strong support in deploying the experiments on IDS Lab's scaled smart city (IDS$^3$C) testbed.

\bibliographystyle{IEEEtran}
\bibliography{ref_drone, IDS_Publications_04022026}
\end{document}